\documentclass[aps,prl,a4paper,twocolumn,noshowpacs,superscriptaddress,floatfix]{revtex4}
\usepackage{graphicx}
\usepackage{amsmath}
\usepackage{amssymb}
\usepackage{graphicx,color}
\usepackage{amsthm}
\usepackage{amsfonts}
\usepackage{algorithmic}
\usepackage{enumerate}
\usepackage{latexsym}
\usepackage{amsmath}
\usepackage{amssymb}
\usepackage[colorlinks=true,citecolor=blue,linkcolor=blue]{hyperref}
\def\avg#1{\left\langle#1\right\rangle}

\begin{document}
\title {Anisotropy engineering edge magnetism in zigzag honeycomb nanoribbons}
\author{Baoyue Li}
\affiliation{School of Physics and Electronic-Electrical Engineering, Ningxia University,
Yinchuan 750021, China}
\author{Yifeng Cao}
\affiliation{Department of Physics, Beijing Normal University,
Beijing 100875, China}
\author{Lin Xu}
\affiliation{Department of Physics, Beijing Normal University,
Beijing 100875, China}
\author{Guang Yang}
\email{yangguang@mail.bnu.edu.cn}
\affiliation{School of Science, Hebei University of Science and Technology, Shijiazhuang, Hebei 050018, China}
\affiliation{Department of Physics, Beijing Normal University,
Beijing 100875, China}
\author{Zhi Ma}
\affiliation{School of Physics and Electronic-Electrical Engineering, Ningxia University,
Yinchuan 750021, China}
\author{Miao Ye}
\affiliation{College of Information Science and Engineering, Guilin University of Technology, Guilin 541004, China}
\author{Tianxing Ma}
\affiliation{Department of Physics, Beijing Normal University,
Beijing 100875, China}

\begin{abstract}
It have been demonstrated that the zigzag honeycomb nanoribbons exhibit an intriguing edge magnetism.
Here the effect of the anisotropy on the edge magnetism in zigzag honeycomb nanoribbons is investigated by using two kinds of large-scale quantum Monte Carlo simulations. 
The anisotropy in zigzag honeycomb nanoribbons is characterized by the ratios of nearest-neighboring hopping integrals $t_{1}$ in one direction and $t_{2}$ in another direction. Considering the electron-electron correlation, it is shown that the edge ferromagnetism could be enhanced greatly as $t_{2}/|t_{1}|$ increase from $1$ to $3$,
which not only presenting the avenue for the control of this magnetism, but also being useful for exploring further novel magnetism in new nano-scale materials.
\end{abstract}
\maketitle
\section{Introduction}
\begin{figure}[htbp]
 \centering
  \includegraphics[width=0.47\textwidth]{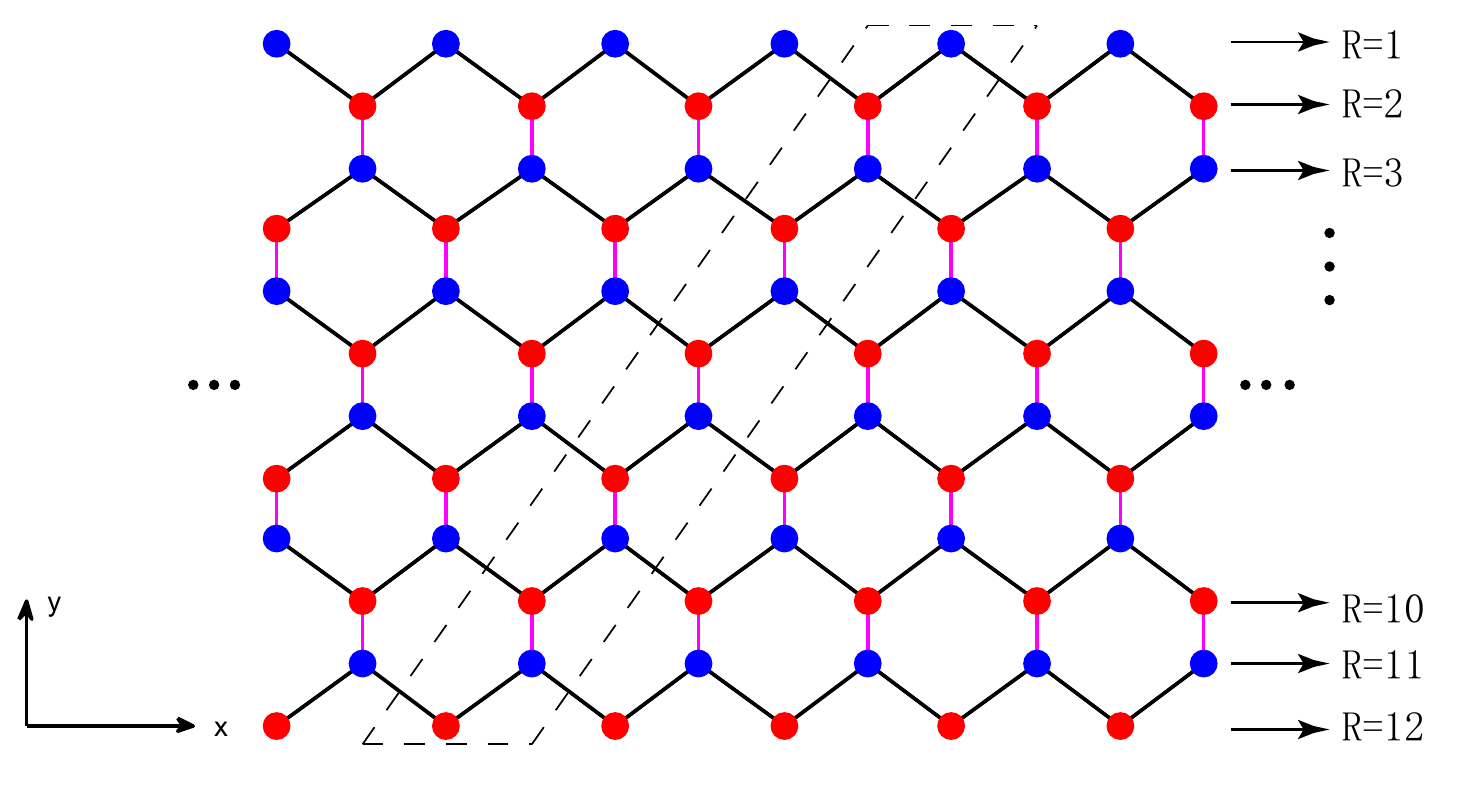}
  \caption{The top view sketch of zigzag honeycomb nanoribbons. The atoms on A(B) sublattices are represented by the blue(red) circles, respectively. The black lines indicate $t_{1}$, and the pink lines indicate $t_{2}$. We adopted the periodic boundary condition in the x-direction and the finite size in the y-direction. The zigzag chains are denoted by index $R$. A unit cell is marked by the dotted line.} \label{fig:lattice}
\end{figure}

Since the discovery of graphene, extensive attention from the research community has been attracted by the emerging honeycomb and honeycomb-like two-dimentional(2D) materials due to their exotic electronic, optical and magnetic properties\cite{Novoselov2012}. The family of these materials includes Hexagonal Boron Nitride, transition-metal dichalcogenides\cite{novoselov2005two,Sun2014}, silicene\cite{gao2013structures,meng2013buckled,yang2018substrate}, germanene\cite{liu2011quantum}, hafnium monolayer\cite{li2013two}, phosphorene \cite{Buscema2014,Castellanos-Gomez2014,Li2014,Liu2014,Xia2014,PhysRevB.89.235319,fei2014strain} as well as its allotropes\cite{guan2014high,zhao2015new,wu2015nine,zhu2014semiconducting,schusteritsch2016single,wang2017psi,zhang2016epitaxial,han2017prediction}, and so forth. As a crucial prerequisite for their practical applications, various methods have been proposed to tailor and generate their properties. Among them, nanopatterning is a fruitful approach because quantum confinement realized in nanostructures often induces strikingly evident quantum phenomina\cite{Oka2014}.
Extensive studies have demonstrated that the local magnetic moments appear on the edge of zigzag graphene nanoribbons(ZGNRs)\cite{son2006half,feldner2011dynamical,cheng2015strain,viana2009magnetism,magda2014gz,PhysRevLett.99.186801},
and the shape of zigzag edge has been shown in Fig.\ref{fig:lattice}, where the top and the bottom of the lattice structure both show the sketch of zigzag edge.
Such quantum phenomenon in ZGNRs instigates more subsequent exploration on the edge magnetism in honeycomb and honeycomb-like nanoribbons such as molybdenum disulfide(MoS$_2$)\cite{Kou2012} and phosphorene\cite{Zhu2014,Du2015,Yang2016}, which may open the avenue to their possible applications in spintronics.
For spintronics, it is required that Curie temperatures of the targeted materials should be higher than ambient temperature,
which is supposed to be approximately room temperature\cite{Yazyev2010}. To solve the challenging problem, further theoretical and experimental investigations are highly demanded.

It has been unveiled that pristine graphene is nonmagnetic due to the vanishing density of states(DOS) at the Dirac point \cite{Kotov2012}. Strikingly, the appearance of edges in a honeycomb-lattice nanostructure gives rise to additional electronic states along the edges at Fermi level which form a quasi-flat band taking up one-third of the one dimensional Brillouin zone in ZGNRs\cite{Fujita1996}. These striking edge states induce the novel magnetic\cite{son2006half,magda2014gz} and optical properties\cite{yang2008magnetic}.
As a well-controlled route, strain engineering is often utilized to modulate the magnetic properties of 2D materials and the corresponding nanostructures \cite{Guinea2010,Johari2012,Ni2008,fei2014strain}. For ZGNRs, applying strain along the zigzag direction have been theoretically proposed to reinforce the edge magnetism\cite{yang2017strain,cheng2015strain,viana2009magnetism}. The anisotropy induced by strain leads to the displacement of the Dirac points. Thus the electronic correlation effect is enhanced by the higher DOS in the extended flat band, which catalyses the enhancement of edge magnetism. The proper strain even could trigger the room-temperature edge magnetism under the suitable Coulomb interaction\cite{yang2017strain}. Distinct from graphene, the puckered structure of phosphorene with a honeycomb lattice endows this material with strong anisotropy\cite{wang2015highly}. Consequently, the quasi-flat band of zigzag phosphorene nanoribbons(ZPNRs) expands across the entire one-dimensional Brillouin zone and it is completely detached from the bulk band\cite{Carvalho2014,Peng2014}. The first-principle and quantum Monte Carlo studies have shown the existence of edge ferromagnetism in ZPNRs which is much more stronger than that in ZGNRs\cite{Zhu2014,Du2015,Yang2016}. Considering the relatively weak Coulomb interaction, it is predicted that the Curie temperature could be even high up to room temperature\cite{Yang2016}. No matter the anisotropy is induced by intentionally introduced strains in the targeted materials or is inborn quality of materials of interest, the study of the anisotropic effect on the ferromagnetism along the zigzag edges of honeycomb nanoribbons
has great academic significance and may advance the development of spintronics.

According to the literature\cite{Lieb1989Two}, honeycomb lattice is bipartite, which can be divided into two sets of sublattices represented by A(blue circle) and B(red circle) in Fig.\ref{fig:lattice}. As is shown in Fig.\ref{fig:lattice}, $t_{1}$ and $t_{2}$ represent two nearest-neighboring hoping integrals and their ratio, namely $t_{2}/|t_{1}|$, denotes the strength of anisotropy, and $t_{2}/|t_{1}|=1.0$ corresponds to the isotropic case of graphene, while for phosphorene, the value of $t_{2}/|t_{1}|$ is near $3.0$. It is interesting to explore the detailed picture of the anisotropy engineering edge magnetism in zigzag honeycomb nanoribbons in the region of $t_{2}/|t_{1}|=1.0\sim 3.0$, which may not only shed more light on some other materials,
but also provide useful information on synthesizing new materials.
In this paper, we use two kinds of large-scale quantum Monte Carlo simulations to explore the anisotropic effect on the edge ferromagnetism of zigzag honeycomb nanoribbons. The edge ferromagnetism is found to be enhanced with the increasing value of $t_{2}/|t_{1}|$ from $1.0$ to $3.0$ under proper interaction because the enhanced interaction effect is caused by the higher DOS located in the extended flat band. Through the picture of the tight-binding model, we found that a band gap show up and becomes broader as $t_{2}/|t_{1}|$ increases. The enhancement of Coulomb interaction and the doping effect on the edge magnetism are also displayed.

\section{Model and Method}

As a prototype of the honeycomb lattice endowed with strong anisotropic nature, phosphorene can be described by a tight-binding model containing five hopping integrals $t_i$(i=1, 2, 3, 4, 5), where $t_{1}=-1.220$ eV, $t_{2}=3.665$ eV, $t_{3}=-0.205$ eV, $t_{4}=-0.105$ eV, $t_{5}=-0.055$ eV\cite{ezawa2014topological}. For ZPNRs, we adopted the periodic boundary along the $x$-direction and finite lattice size in the $y$-direction. It has been verified that the anisotropic effect of ZPNRs on edge magnetism is mainly reflected by the nearest hopping terms $t_{1,2}$ due to their much higher values than those of $t_{3,4,5}$\cite{ezawa2014topological}. Therefore, here our study focuses on the correlation of edge magnetism and $t_{2}$ to $|t_{1}|$ ratios with the vanishing $t_{3,4,5}$ under Coulomb interaction in the honeycomb nanoribbons.

The single-band Hubbard model is employed to describe the honeycomb nanoribbons and the Hamiltonian is given as
\begin{equation}
H=\sum_{<ij>}{t_{ij}}c_{i\sigma }^{\dagger}c_{j\sigma }+U\sum_{i}n_{i\uparrow}n_{i\downarrow}-\mu\sum_{<i>}c_{i\sigma }^{\dagger}c_{i\sigma }\
\end{equation}
where $t_{ij}$ represents the hopping integral between the $i$-th and $j$-th sites and we consider $t_{2}/|t_{1}|=1.0, 2.0, 3.0$ to explore the anisotropic effect on the edge magnetism. $c_{i\sigma }(c_{i\sigma}^{\dagger})$ denotes the annihilation (creation) operator of electron at the $i$-th site and $n_{i\sigma}=c_{i\sigma}^{\dagger}c_{i\sigma}$ is the occupation number operator. $\mu$ is the chemical potential and $U$ is the on-site Coulomb repulsion. As the powerful tools for treating the strong correlated systems, the determinant quantum Monte Carlo (DQMC)\cite{hirsch1985two,blankenbecler1981monte,santos2003introduction} and the constrained path quantum Monte Carlo (CPQMC) methods\cite{zhang1995s} are utilized to simulate magnetic correlation in the presence of Coulomb interaction\cite{ma2013quantum,PhysRevLett.120.116601,ma2014possible,ma2011pairing,wu2013ground,ma2015triplet}. The results of DQMC can exhibit the properties of the related systems at finite temperature, while the CPQMC is designed to explore the ground-state properties. For the DQMC, it is free from the notorious sign problem in the half filled cases due to the particle-hole symmetry, which we mainly care about here and thus the corresponding results are guaranteed to be reliable\cite{Yang2016}. To explore the effect of electron fillings, we present some results which are very near to the half filling
by using the CPQMC, and CPQMC is a method inborn to avoid the sign problem.

To explore the thermodynamic properties of the edge magnetism in honeycomb nanoribbons, the uniform magnetic susceptibility $\chi$ along each edge at finite temperatures is calculated using the DQMC. The uniform magnetic susceptibility is defined as the zero-frequency spin susceptibility in the $z$ direction as
\begin{equation}
\chi=\int_{0}^{\beta }\mathrm{d}\tau \sum_{ij}\langle S_{i}(\tau)\cdot S_{j}(0)\rangle
\end{equation}
where $S_i(\tau)=e^{H\tau}S_i(0)e^{-H\tau}(\hbar=1)$ with $S_{i}=c_{i\uparrow}^{\dagger}c_{i\uparrow}-c_{i\downarrow}^{\dagger}c_{i\downarrow}$ . At first, summation run over the sites along each edge, and then the edge magnetic susceptibility is obtained through averaging the results of the top edge and the bottom edge. Furthermore, the spatial distribution of the magnetic correlations is elucidated utilizing the CPQMC method to calculate the equal-time magnetic structure factor for each zigzag chain which is defined as
\begin{equation}
M_{R}=\frac{1}{L_{x}^{2}}\sum_{i,j\in Row}S_{i,j}
\end{equation}
where $S_{i,j}= \langle S_{i}\cdot S_{j}\rangle$. R is the index of the zigzag chain, i,j are the indices of the sites along the R-th zigzag chain, and $L_x$ represents the number of sites in each zigzag chain. $M_{R}$ is calculated along the zigzag chain from the bottom to the top as shown in Fig.\ref{fig:lattice}. Through the values of spin structure factor $M_{R}$, the spatial distribution of spin correlations could be clearly presented.

\section{Results and Discussion}
\begin{figure}[bpt]
 \centering
  \includegraphics[width=0.425\textwidth]{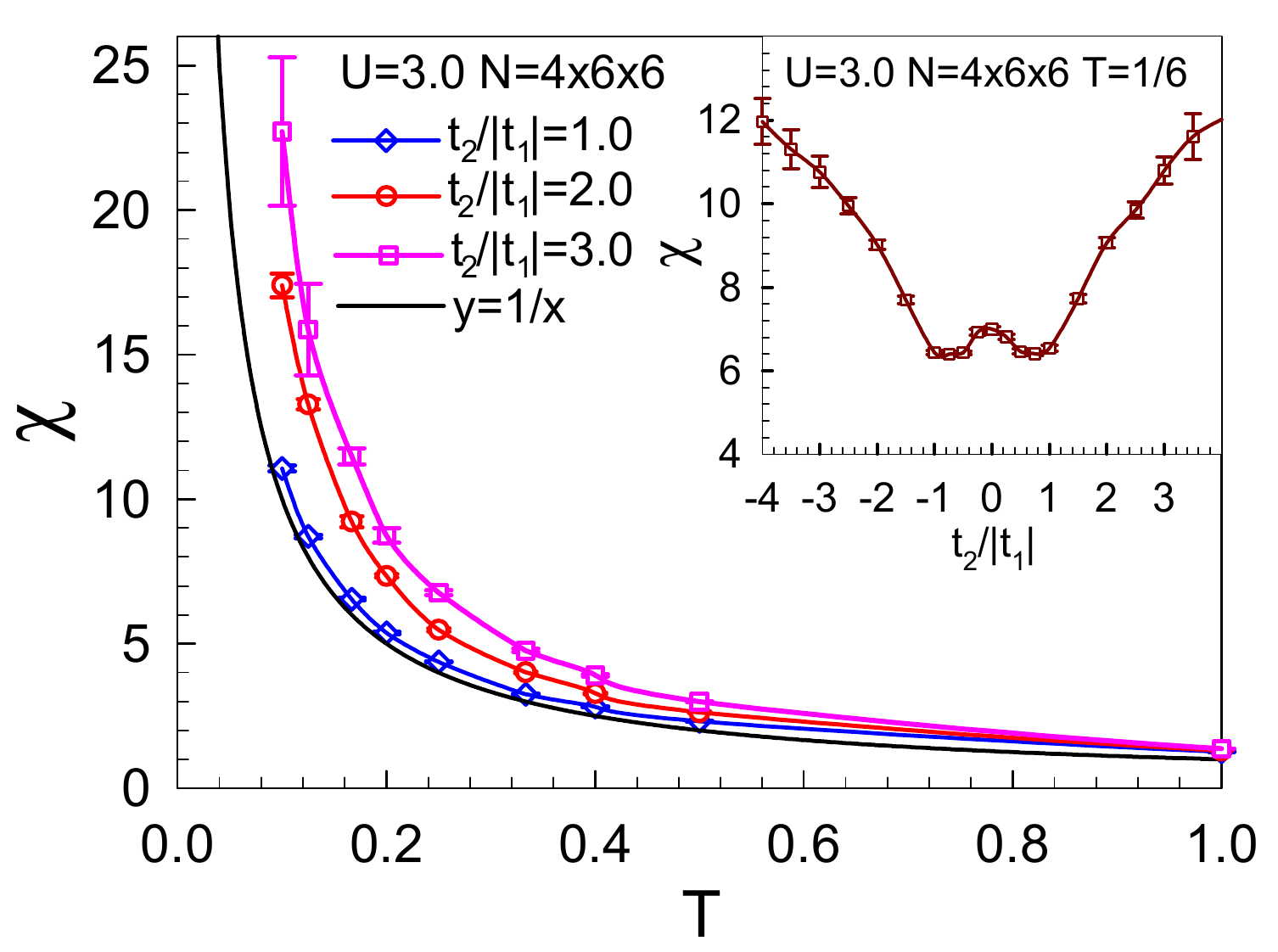}
  \caption{The edge magnetic susceptibility dependent on the temperature with different $t_{2}$/$|t_{1}|$ at half filling, $U=3.0$ and $N=4\times 6\times 6$. Inset:The edge magnetism as a function of $t_{2}$/$|t_{1}|$ with the certain temperature $T=1/6$ at half filling, $U=3.0$ and $N=4\times 6\times 6$. }\label{fig:anisotropy}
\end{figure}

To shed light on the anisotropic effect on the edge magnetism in the zigzag honeycomb nanoribbons, the Fig.\ref{fig:anisotropy} is plotted to exhibit the magnetic susceptibility along the zigzag edge as a function of temperature with different ratios of $t_{2}$ to $|t_{1}|$ at half filling, Coulomb interaction $U=3.0$ and lattice size $4\times 6\times 6$. In the following, we take $|t_1|$ as the unit if there is no special illustration. For graphene-based material, $|t_1|$ is around 2.7 eV, and for phosphorene, $|t_1|$ is around 1.220 eV. The value of the on-site repulsion $U$ can be taken from its
estimation in polyacetylene\cite{RevModPhys.81.109,PhysRevLett.97.146401,doi:10.1063/1.1747540} $U\cong$6.0-17 eV,
which clearly spans a large range of values for graphene based materials, and latter Peierls-Feynman-Bogoliubov variational principle shows
that $U\simeq$ 4 eV is reasonable for graphene, silicene, and benzene\cite{PhysRevLett.111.036601}.
Therefore, to explore the importance of interactions on the magnetism of nanoribbons under study,
we study the model Hamitonian in the range of $U/|t_1|$=$1\sim 5$, and this is also feasible for phosphorene\cite{Du2015,Yang2016}.

Apparently, the correlations of edge magnetic susceptibility and temperature display the Curie-Weiss behavior $\chi=A/(T-T_{c})$ which describes the magnetic susceptibility $\chi$ dependent on the temperature above the Curie temperature $T_{c}$. According to the reference line $y=1/x$, all the lines for $t_{2}/|t_{1}|=1.0,2.0,3.0$ diverge at the finite low temperature with $U=3.0$ suggesting that the zigzag honeycomb nanoribbons have the ferromagnetic behavior. Moreover, $\chi$ increases with the increasing $t_{2}/|t_{1}|$ at low temperature which presents the enhancement of the anisotropy for the edge magnetism in zigzag honeycomb nanoribbons. To provide clearer diagram for the correlation between $\chi$ and $t_{2}/|t_{1}|$, an inset is added in Fig.\ref{fig:anisotropy}. When the absolute value of $t_{2}/|t_{1}|$ is larger than $1.0$ up to $4.0$, the edge magnetic susceptibility almost linearly increases with the increasing $t_{2}/|t_{1}|$ as is shown in the inset of Fig.\ref{fig:anisotropy}. While, for the absolute value of $t_{2}/|t_{1}|$ smaller than $1$ down to $0$, the magnetic susceptibility slightly increase, which is similar as that in zigzag graphene nanoribbons\cite{yang2017strain}. Therefore, we may assert that stronger anisotropy can induce stronger edge magnetism in zigzag honeycomb nanoribbons.

\begin{figure}[bpt]
 \centering
  \includegraphics[width=0.425\textwidth]{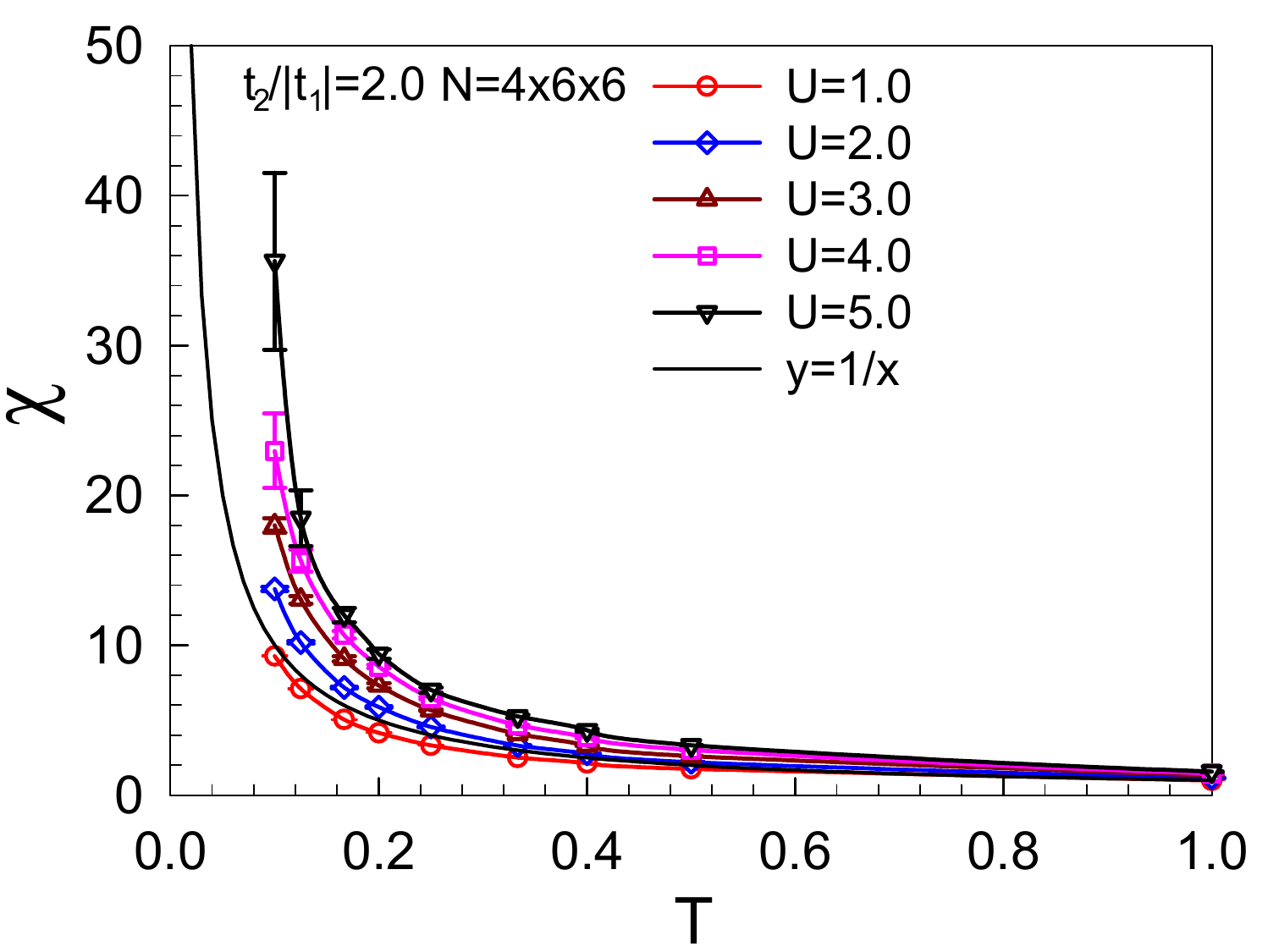}
  \caption{Temperature dependence of magnetic susceptibility $\chi$ for the different Coulomb interaction with the certain $t_{2}/|t_{1}|=2.0$ at half filling and $N=4\times 6\times 6$.}\label{fig:interaction}
\end{figure}

To understand the physical scenarios induced by Coulomb interaction $U$, the magnetic susceptibility $\chi$ of zigzag honeycomb nanoribbons with different Coulomb interactions $U$ is computed at the same $t_{2}/|t_{1}|$ as Fig.\ref{fig:interaction} illustrates. Clearly, $\chi$ is enhanced by the interaction $U$ at the same temperature and $t_{2}/|t_{1}|$. In addition, the system is dominated by the ferromagnetic fluctuation at $U\geq 2.0$ and $t_{2}/|t_{1}|=2.0$. Hence, Fig.\ref{fig:anisotropy} and Fig.\ref{fig:interaction} show that both anisotropy and interaction can make the edge ferromagnetism robust in the zigzag honeycomb nanoribbons.

\begin{figure}[bpt]
 \centering
  \includegraphics[width=0.425\textwidth]{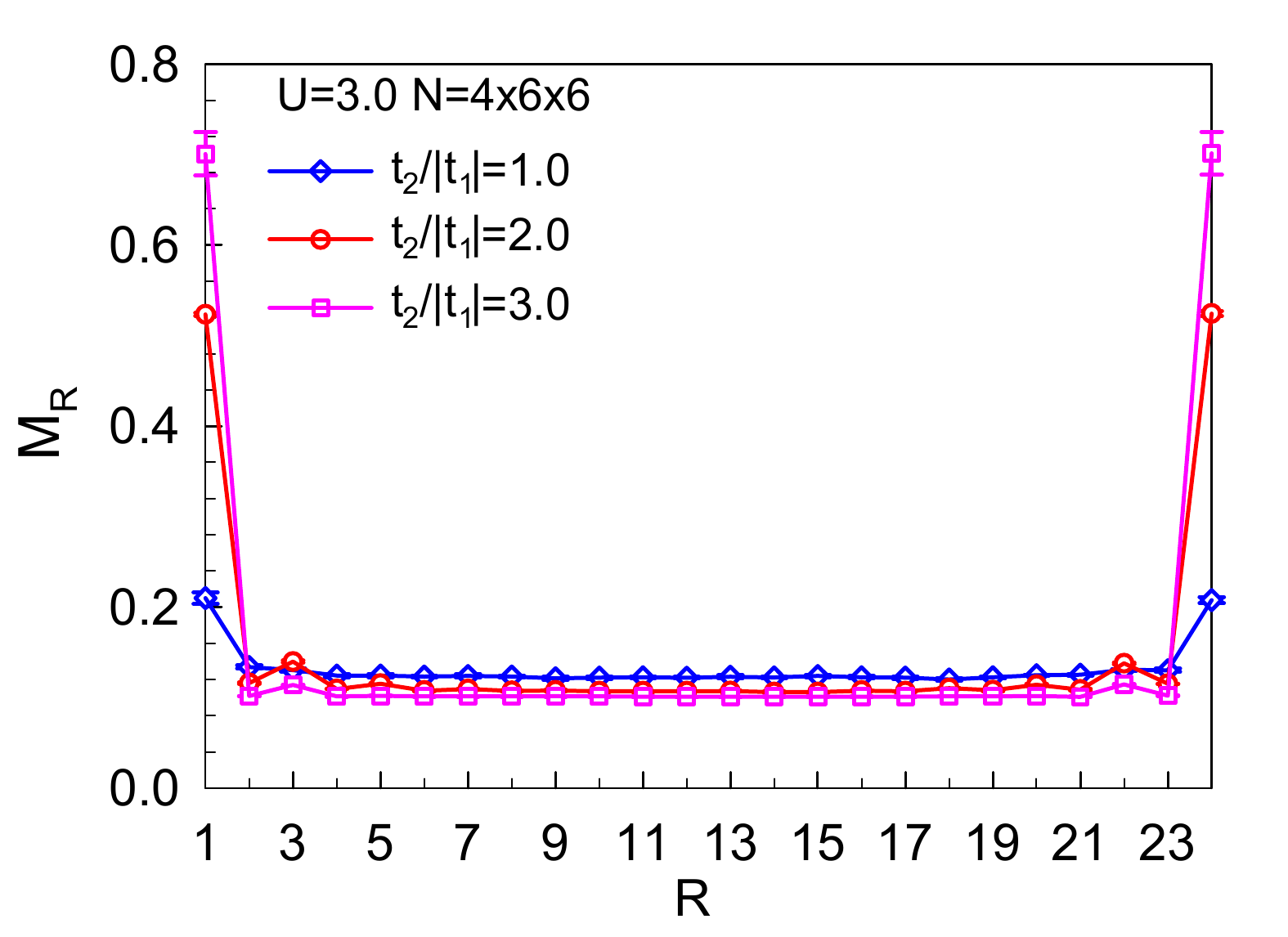}
  \caption{The magnetic structure factor for each row with different $t_{2}/|t_{1}$ at half filling, $U=3.0$ and $N=4\times 6\times 6$.}\label{fig:anisotropy_cpmc}
\end{figure}

\begin{figure}[bpt]
 \centering
  \includegraphics[width=0.425\textwidth]{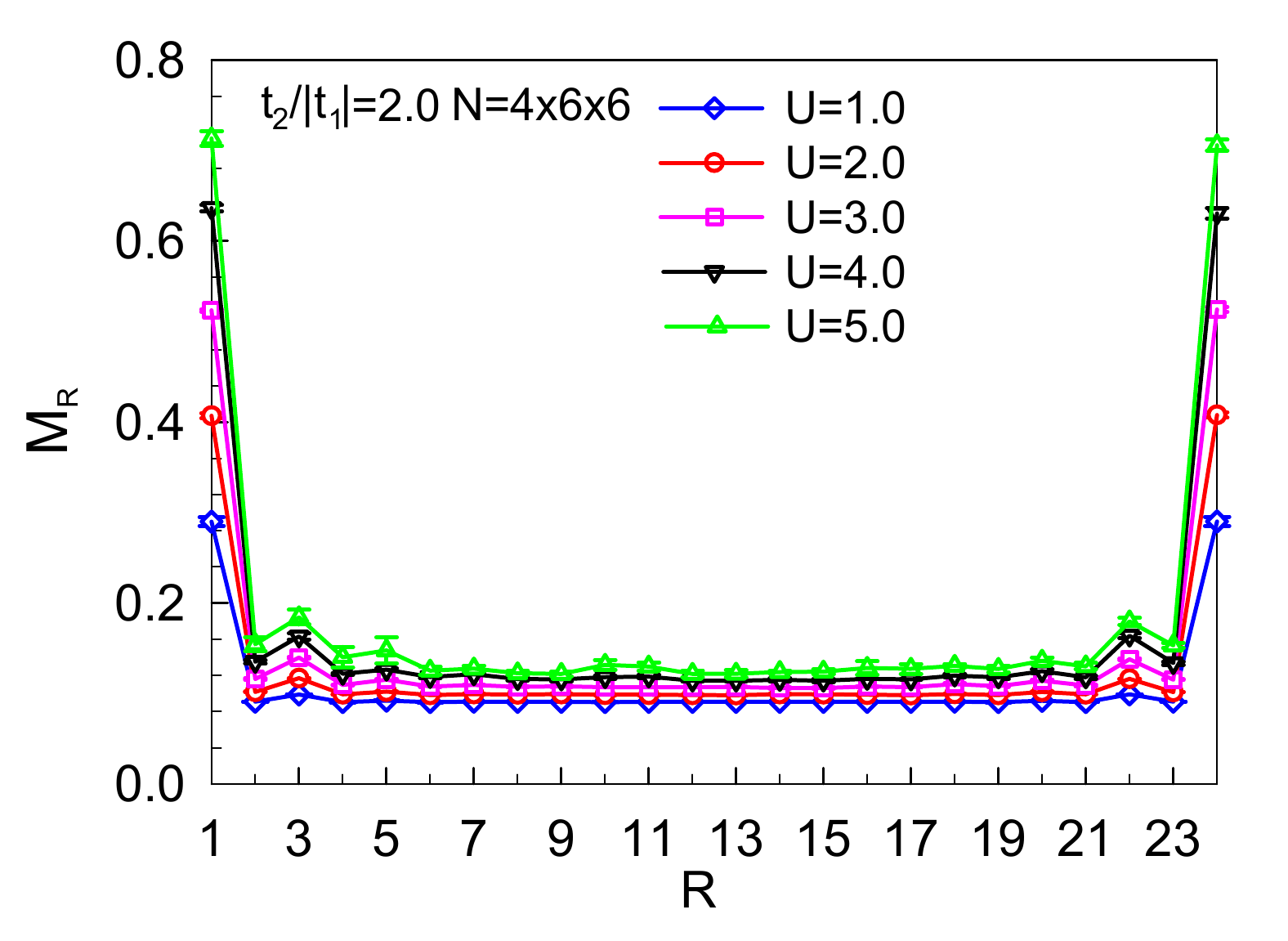}
  \caption{The magnetic structure factor for each row with different Coulomb interactions at half filling, $t_{2}/|t_{1}|=2.0$ and $N=4\times 6\times 6$.}\label{fig:interaction_cpmc}
\end{figure}

To further study the spatial distribution of magnetic correlations, CPQMC is used to calculate the equal-time magnetic structure factor $M_{R}$ along each zigzag chain. Fig.\ref{fig:anisotropy_cpmc} presents $M_{R}$ with different cases of $t_{2}$/$|t_{1}|=1.0,2.0,3.0$ at $U=3.0$ , half filling and $N=4\times 6\times 6$.
For an half-filled Hubbard model on a perfect honeycomb lattice, the system
shows antiferromagnetic correlations\cite{PhysRevLett.120.116601}. As the structure of the
honeycomb lattice can be described by two interpenetrating
sublattices, the spin correlation between the nearest-neighbor
sites (or sites on different sublattice) is negative, due to antiferromagnetic correlations,
while the spin correlation between the sites belonging to the same sublattice,
for example, between the next nearest-neighbor sites, has to be positive.
The $M_{R}$ defined here is an average of the spin correlation between sites belonging to the same sublattice, thus it is positive and acts like ferromagnetic behavior\cite{cheng2015strain}.

 The value of $M_{R}$ is dramatically larger along each edge than that along each chain in the bulk so that the magnetic correlations are mainly distributed along each edge. Meanwhile, we can see that the edge magnetic correlations become larger with the increasing values of $t_{2}$/$|t_{1}|$. Thus the enhancement of the anisotropy for edge magnetism is further verified by the results of the CPQMC in agreement with the conclusion obtained from the DQMC. In Fig.\ref{fig:interaction_cpmc}, the results of the CPQMC illustrate $M_{R}$ at each chain dependent on the Coulomb interactions at the same $t_{2}$/$|t_{1}|$. It is clear that the larger interaction leads to the stronger edge magnetism which is also consistent with the results of DQMC. Even the magnetic structure factor has a finite positive value at $U=1.0$, that does not mean the exact presence of observed magnetism,
 for which we have to make careful finite size scaling analysis to explore the properties at thermodynamical limits.
 This does cost huge CPU time and restrict us. Anyway, the results shown in Fig.\ref{fig:interaction_cpmc} at least demonstrate that the magnetic structure factor is enhanced greatly as the interaction strength increases.

\begin{figure}[bpt]
 \centering
  \includegraphics[width=0.425\textwidth]{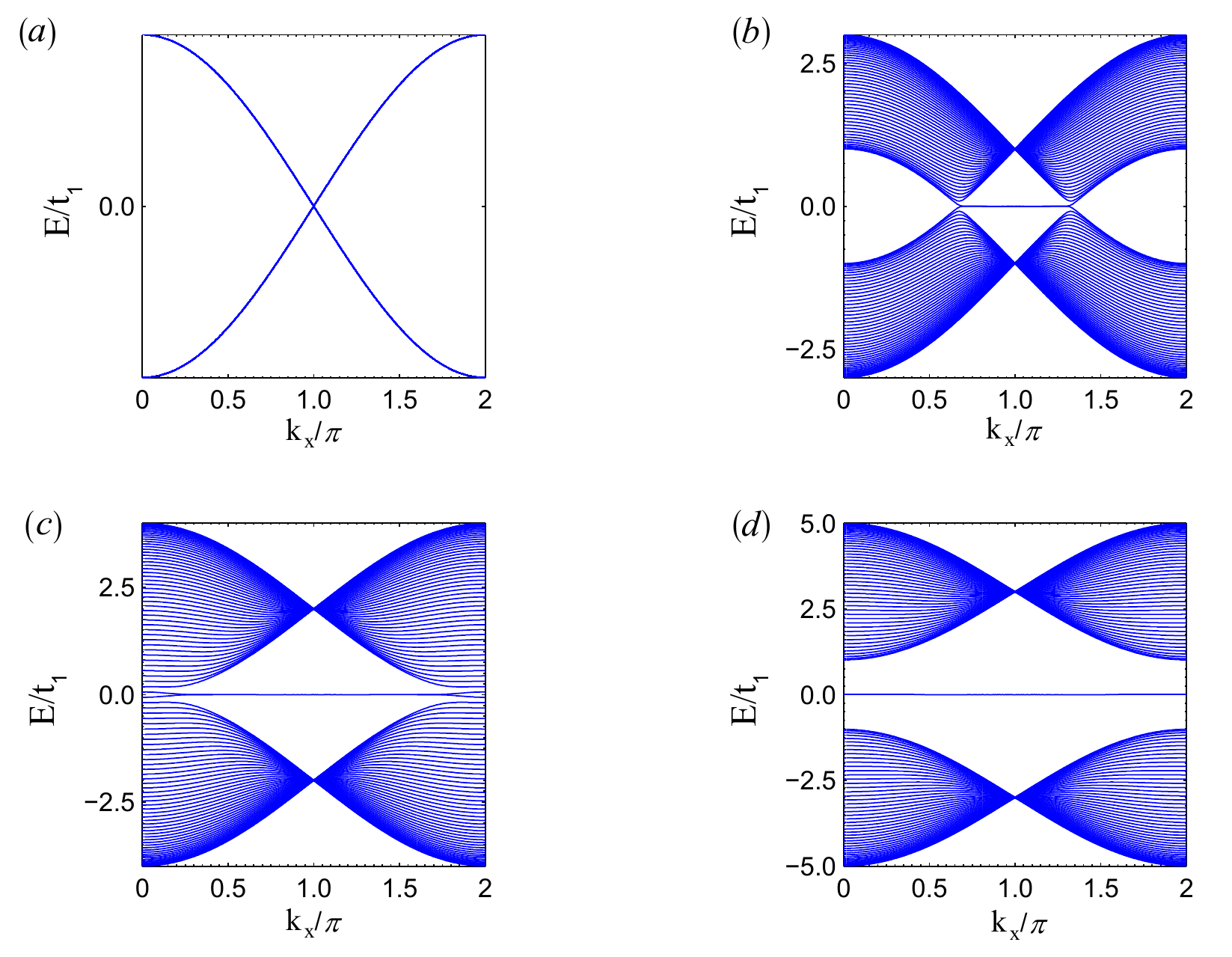}
  \caption{The band structure of the zigzag honeycomg nanoribbons with (a)$t_2/|t_{1}|=0.0$, (b)$t_2/|t_{1}|=1.0$, (c)$t_2/|t_{1}|=2.0$, (d)$t_2/|t_{1}|=3.0$. }\label{fig:bandstructure}
\end{figure}

The variation of the topology of the band structure caused by $t_{2}/|t_{1}|$ reveals the nature of the enhanced edge magnetism induced by the anisotropy in such systems as is presented in Fig.\ref{fig:bandstructure}. For the case of $t_{1}=-1.0$ and $t_{2}=1.0$ in Fig.\ref{fig:bandstructure}(b), the band structure corresponds to that of zigzag graphene nanoribbons with two Dirac cones at $K$ and $K^{\prime}$. A flat band consisting of the edge states connects these two Dirac points. The flat band takes up one-third of the one dimensional Brillouin zone. We take $t_{1}$ as the unit and increase $t_{2}$. As the $t_{2}=2.0$ corresponds to $t_{2}/|t_{1}|=2.0$ in Fig.\ref{fig:bandstructure}(c), we can see that two Dirac cones approach to $\Gamma(k=0)$ and then the flat band extends dramatically. In Fig.\ref{fig:bandstructure}(d), we set $t_{2}=3.0$ and $t_{1}$ as the unit, and $t_{2}/|t_{1}|$ is equivalent to $3.0$ which approximately corresponds to zigzag phosphorene nanoribbons according to Ref.\cite{ezawa2014topological}. In this condition, Fig.\ref{fig:bandstructure}(d) shows a flat band occupying the entire one dimensional Brillouin zone. In the meanwhile, a band gap opens up in the bulk with the increasing anisotropy. The extended flat band derived from the increasing $t_{2}/|t_{1}|$ leads to the higher density of states at Fermi level which enhances the interaction effect. Thereby the stronger ferromagnetism is induced by the stronger anisotropy.

\begin{figure}[bpt]
 \centering
  \includegraphics[width=0.425\textwidth]{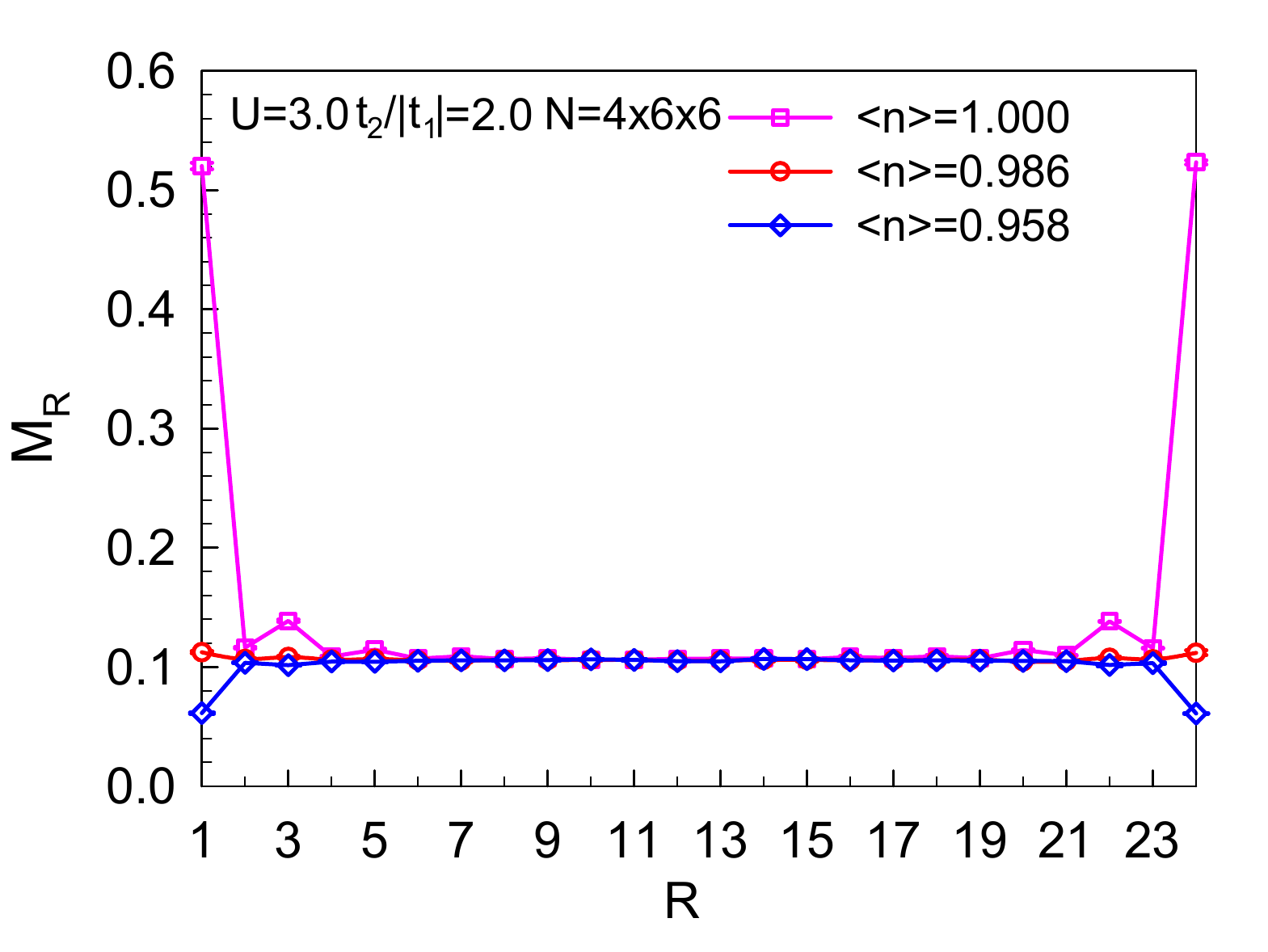}
  \caption{The magnetic structure factor for each row at different electron fillings with $U=3.0$, $t_{2}/|t_{1}|=2.0$ and $N=4\times 6 \times 6$.}\label{fig:doping_cpmc}
\end{figure}

Finally, the doping effect on the edge magnetism is explored using the CPQMC. The relation between the magnetic structure factor and the electron filling $\avg{n}$ is illustrated in Fig.\ref{fig:doping_cpmc}. It is clear that the edge ferromagnetism is sharply weakened as the electron filling moves away from the half filling and the doped charge mostly locates along the edge. Therefore, it may give a possible way to manipulate the edge magnetism in the honeycomb nanoribbons. The doping level presents in Fig. \ref{fig:doping_cpmc} is $\delta=1-\avg{n}$=0.014 and 0.042 respectively, namely, 1.4 percent or 4.2 percent doping ratio, which are within the current experimental capacity, as in graphene and other honeycomb-like 2D materials,  doping achievable by gate voltage or chemical doping is usually on the order of $10^{12}\sim10^{13}$ cm$^{-2}$\cite{Novoselov2012}.

\section{Summary}
In summary, we used both the DQMC and CPQMC methods to explore the effect of the anisotropy, the interaction and the doping on the edge ferromagnetism in the honeycomb nanoribbons. At a fixed Coulomb interaction, for example $U=3.0$, which is a reasonable interaction strength for various 2D materials with honeycomb-like structure, our intensive numerical results show that the edge magnetism could be enhanced remarkably as $t_{2}/|t_{1}|$ increases from $1$ to $3$.
For a fix $t_{2}/|t_{1}|=2.0$, a ferromagnetic-like behavior is predicted as $U\geq 2.0$, and the ferromagnetic correlation is reduced greatly with a finite doping.
These results provide a route for tailoring the magnetic properties of honeycomb 2D materials and searching for new materials with the honeycomb lattice.
\section*{Acknowledgements}
This work was supported by NSFC (No. 11774033) and Beijing Natural Science Foundation (No.1192011).
We also acknowledge computational support from HSCC of Beijing Normal University.
\bibliography{reference}
\end{document}